\documentclass[dvips]{article}
\newcommand{\bm}[1]{\mbox{\boldmath$#1$}}
\usepackage{amsmath}
\usepackage{amssymb}
\usepackage{latexsym}
\usepackage{graphics}
\usepackage{graphicx}
\def\be{\begin{equation}}
\def\ee{\end{equation}}
\def\bea{\begin{eqnarray}}
\def\eea{\end{eqnarray}}
\def\esp{\end{split}}
\def\bsp{\begin{split}}
\def\nn{\nonumber}
\begin{document}
\title{Spectral approach to the inverse problem for the field of arbitrary changing electric dipole}
\author{V. Epp\thanks{Corresponding author. Email: epp@tspu.edu.ru}  and J. G. Janz\\
\vspace{6pt} {\em Tomsk State Pedagogical University, Tomsk, Russia}
}
\date{}
\maketitle
\begin{abstract}
The inverse problem for electromagnetic field produced by arbitrary altered charge distribution in dipole approximation is solved. The charge distribution is represented by its dipole moment. It is assumed that the spectral properties of magnetic field of the dipole are known. The position of the dipole and its Fourier components are considered as the unknown quantities.  It is assumed that relative increments of amplitude and phase of magnetic field in the vicinity of the observation point are known.  The derived results can be used for study of phenomena concerned with occurrence and variation of localized electric charge distribution, when the position and the dynamics of a localized source of electromagnetic field are to be defined.

\vspace{1em}

\end{abstract}
%

\section{Introduction}
The early studies of the inverse problem in electrodynamics have been devoted to the inverse scattering problem for electromagnetic waves. 
In succeeding years the focus was turned to the theory of  inverse boundary values problem, see for examples  \cite{Ola, Chadan, Ola2003}.  A large variety of papers is devoted to inverse problem for the field of static magnetic or static electric dipole in connection with medical imaging. For example, the current sources in the brain produce external magnetic fields and scalp surface potentials that can be measured using magnetoencephalography and electroencephalography (see for examples \cite{Mosher, Erkki}). The solution in a particular case of a static  dipole field was obtained in  \cite{EppDip}. 

Special field of research is constituted by  the inverse problems for electromagnetic field of a point-like source like a moving charge or varying dipole. There was a number  of papers which can be considered as devoted to the inverse problem of electromagnetic radiation of a charge moving in bending magnets and undulators.  
For example, some ways of solution of this problem was suggested in  \cite{ bagrovNIM}.
Solution of the inverse problem of  radiation of a moving point-like charge in a far-field approximation was presented in  \cite{bagrov85}. 
 A general solution of the inverse problem for the Li\'{e}nard-Wiechert potentials of an arbitrary moving charged particle was given in  \cite{epp04}.

Another elementary and widespread source of electromagnetic field is a variable electric or magnetic dipole. 
The electromagnetic field of an arbitrary distribution of charge at the distance much greater than the characteristic dimension of the area where the charge is distributed can be described in a certain approximation as a field of a point-like dipole.
There are many phenomena in which electromagnetic field of localized distribution of charge is known or can be measured, and it is necessary to  restore the properties of the source. For example, investigations of electric charge distribution in the thundery clouds, or emission of electromagnetic waves at the boundary of continental platforms which can be used for earthquakes forecasting (see for example, \cite{Fraser, Mogna}),
or investigation of development of cracks in crystals \cite{inv3, inv3-1, inv3-3} (see also recent discussion on  identification of the position and shape of cracks detected in elastic solid samples \cite{Brigante}).

  In this paper we  assume that the source of the field is localized in an area which is much smaller than the distance between the source and the observer and can be replaced by an arbitrary varying dipole.  This does not mean that we consider only the so-called far zone, because the far zone is defined as area at distances much greater than the wave length. But a non relativistic charged particle (which is the case of this paper) generates electromagnetic field with characteristic wave length much greater than the area in which the particle moves. Hence, we consider here both, the near field zone  within the distance comparable or  less then the wave length and the far field zone.

 In our previous paper  \cite{Inver1} we have solved the inversion problem for the field of an arbitrary changing electric dipole. It was assumed that at some point of the space the electric and magnetic fields of a dipole are known as the functions of time. As a result of the solution, formulae were derived for  the field source position and for the dynamics of the dipole. Such approach is not very convenient for practical applications, because the  measurements of fast changing electromagnetic fields as the functions of time face some difficulties. Usually, it is much easier to measure the amplitude of electric or magnetic field at some frequency of spectrum. From this point of view it can be of interest such definition of the inverse problem when the given quantities are the spectral characteristics of the field, and  the desired solution is the position and spectral properties of the field source.


As we shall see further, the vector of magnetic field of the dipole is orthogonal to the line connecting the dipole with an observer. Hence, if we define the harmonic component of magnetic field vector, we can define the line between the source and the observer. In this paper we present a solution of the inverse problem for the case when the spectral components of magnetic field at some point are given or we know the gradient of the spectral components at the point of observation. It is shown that the information  only on the spectral properties of the magnetic field is not sufficient for solution of the inverse problem. But if the gradient of the field at the observation point is known, the full solution of the problem is possible.
%
\section{Inverse problem for spectral components of the magnetic field}
%
Suppose we know the magnetic field which is produced by some charge distribution  at fixed frequency $\omega$ at some point of space. Suppose the distance between this point and some point inside the area where the charge is distributed is much greater than the dimensions of the area. In this case we can replace the charge distribution by its electric dipole. Then the problem is to find the position of  the dipole and the Fourier amplitude of the dipole. 

\begin{figure}[htbp]\center
\includegraphics[width=1.9in]{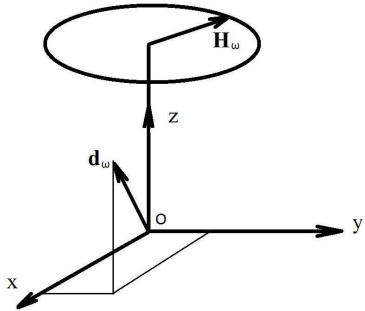}
\caption{Reference system.}
\label{fig_0}
\end{figure}

Fourier transform of the magnetic field is given by  \cite[\S 72]{Landau} (Gaussian system of units is used):
\be
\label{spectr}
\bm H_\omega (\bm r)=\frac{{\rm i} k}{r^2}(\bm d_\omega\times \bm n)({\rm i} kr-1){\rm e}^{{\rm i} kr},
\ee
where $\bm d_\omega$ is the Fourier transform of the electric dipole, $\bm r$ is the radius-vector of the observation point, $\bm n =\bm r/r$ is the unit vector pointing from the dipole to the observer, $k=\omega/c$ is module of the wave vector. In this section we assume that the magnetic field is not linearly polarized. Then we can define the plane to which vector $\bm H$ is bound. Let the axis $OZ$ be orthogonal to this plane
 as shown in Fig. \ref{fig_0}.
  Then the harmonic component of the magnetic field vector will describe an ellipse in the plane $XY$, which we call polarization ellipse of vector  $\bm H_\omega$. 
Now we know that the source of the field is on $OZ$-axis, but we do not know in which direction of the axis -- positive or negative it lies. If the magnetic field is linearly polarized, one cannot define the axis $OZ$ unambiguously.  Next, we represent vector $\bm d_\omega$ by its Cartesian coordinates
\be\label{dcomp}
\bm d_\omega=(d_{0 x}\exp {\rm i} \alpha_x,\, d_{0 y}\exp {\rm i} \alpha_y,\,d_{0 z}\exp {\rm i} \alpha_z),
\ee
where $\alpha_j$ are the initial phases, $d_{0j}$ are the real positive amplitudes.

Respectively, Cartesian coordinates of vector $\bm H_\omega$ are
\be\label{Homega}
\bsp
H_{\omega x}&=\frac{{\rm i} k}{r^2}\varepsilon d_{0 y}({\rm i} kr-1)\exp[{\rm i} (kr+\alpha_y)], \\
H_{\omega y}&=-\frac{{\rm i} k}{r^2}\varepsilon d_{0 x}({\rm i} kr-1)\exp[{\rm i} (kr+\alpha_x)],
\end{split}
\ee
where $\varepsilon=n_z=\pm 1$.

The size and orientation of the magnetic field polarization ellipse can be defined in terms of polarization parameters, for example, Stokes parameters. Polarization parameters are quadric combinations of components  $H_{\omega x}$ and $H_{\omega y}$ and can be easily measured experimentally. We define the Stokes parameters as in  \cite{Born}:
\bea\label{Stokes1}
\bsp
&s_0=H_{\omega x}H^*_{\omega x}+H_{\omega y}H^*_{\omega y},\\
&s_1=H_{\omega x}H^*_{\omega x}-H_{\omega y}H^*_{\omega y}, \\
&s_2=H_{\omega x}H^*_{\omega y}+H_{\omega y}H^*_{\omega x},\\
&s_3={\rm i} (H_{\omega y}H^*_{\omega x}-H_{\omega x}H^*_{\omega y}).
\end{split}
\eea
We would like to  emphasize that the field  (\ref{Homega}) is not the field of plane wave. It means that you cannot calculate Stokes parameters from the electric field, as it is usually done. The electric field is neither equal nor orthogonal to the  magnetic one.
Evidently, one can represent the Stokes parameters in the form
\bea\label{Stokes2}
\bsp
s_0=&|H_{\omega x}|^2+|H_{\omega y}|^2, \\
s_1=&|H_{\omega x}|^2-|H_{\omega y}|^2, \\
s_2=&2|H_{\omega x}||H_{\omega y}|\cos (\phi_x-\phi_y), \\
s_3=&2|H_{\omega x}||H_{\omega y}|\sin(\phi_x-\phi_y),
\end{split}
\eea
where  $\phi_x-\phi_y$ is the phase difference between the projections of  $\bm H_\omega$ onto axes $OX$ and $OY$.

Substituting expressions (\ref{Homega}) into (\ref{Stokes1}), we get:
 \bea\label{Stokes3}
 \bsp
&s_0=J(d_{0 x}^2+d_{0 y}^2), \\
&s_1=J(d_{0 y}^2-d_{0 x}^2), \\
&s_2=-2Jd_{0 x}d_{0 y}\cos (\alpha_x-\alpha_y),\\
&s_3=2Jd_{0 x}d_{0 y}\sin (\alpha_x-\alpha_y).
\end{split}
\eea
   Here
   \be\label{mashtab}
   J=\frac{k^2(1+k^2r^2)}{r^4}.
\ee 
    Comparing (\ref{Stokes2}) and (\ref{Stokes3}), we see that projection of the dipole polarization ellipse onto plane  $XY$ is equal to that of magnetic field but is turned by angle  $\pi/2$.  It follows also from  (\ref{Stokes3})  that if the Stokes parameters are known, then the dipole components can be defined only up to a scaling multiplier. Let us denote $\delta_i=\sqrt{I}d_{0 i}$. Then the solution of  (\ref{Stokes3}) takes the form
\bea\label{delta}
\delta_x=\sqrt{\frac{s_0-s_1}{2}},\quad 
\delta_y=\sqrt{\frac{s_0+s_1}{2}}, \\
\cos (\alpha_x-\alpha_y)=-\frac{s_2}{\sqrt{s_0^2-s_1^2}},\nn\\
\sin (\alpha_x-\alpha_y)=\frac{s_3}{\sqrt{s_0^2-s_1^2}}.\nn
\eea      
These relations define the dipole polarization ellipse up to a scaling multiplier.
  
   Hence, knowing the spectral properties or polarization of the dipole magnetic field we can define the line connecting dipole with the observer, and projection of the dipole polarization ellipse onto the plane orthogonal to that line. Still remain unknown the absolute value of the dipole, one of the two directions to the field source, and the distance to the source. Some of these quantities can be find if the spectral components of the electric field are known. In this paper we discuss another possibility to introduce additional input data. Namely, we suppose that the gradient of spectral amplitude of the magnetic field is known. We start with the  most general case -- we suppose that all derivatives of $\bm H_\omega$ with respect to all coordinates are known. In Section \ref{sec4} we present some solutions of the inverse problem  when not all the derivatives are used.
\section{Inverse problem for derivatives of the magnetic field}        \label{sec3}
            Let us rewrite coordinates of vector $\bm H_\omega$  in an arbitrary Cartesian coordinate system $(x_1,x_2,x_3)$ as follows 
 $H_{\omega m}=|H_{\omega m}|\exp {\rm i} \varphi_m$, $m=1,\,2,\, 3. $
Next, we take derivative over coordinate  $x_l$ and divide the result by  $H_{\omega m}$. This produces a matrix
\be
\label{MatD}
\Delta_{lm}\equiv\frac{1}{H_{\omega m}}\frac{\partial H_{\omega m}}{\partial x_l}=\frac{1}{|H_{\omega m}|}\frac{\partial |H_{\omega m}|}{\partial x_l}+{\rm i}\frac{\partial \varphi_m}{\partial x_l}.
\ee             
As we see, the real part of the matrix element represents relative  rate of change of the magnetic field amplitude along the respective axis. Imaginary part of   $\Delta_{lm}$ is the phase rate.
   
Suppose all matrix elements $\Delta_{lm}$  are known. The real part of each element can be measured as ratio of the magnetic field  increment to coordinate increment for two near points of space. Respectively, the imaginary part can be measured as relative increment of phase for the same two points. 

Now the inverse problem is defined as follows: there are 9 complex or 18 real numbers  $\Delta_{lm}$ (because each complex number is defined by two real numbers). Two vectors have to be found -- a real vector $\bm r$, and a complex vector $\bm d_\omega$. This makes 9 unknown real quantities. Hence, the problem is  over-specified, which means that the matrix elements  $\Delta_{lm}$ are not independent -- there are some constraint equations which we shall ascertain later on.

  Let us take derivatives from (\ref{spectr}) with respect to coordinates and substitute the results in  (\ref{MatD}). We introduce a dimension\-less matrix  $D_{lm}=\Delta_{lm}/k$. The matrix elements are:
 \bea
  \label{matrix}
 D_{lm}=
  \left(\begin{array}{ccc}   
     0 & p_2 d_{\omega 3}& -  p_3 d_{\omega 2}\\
     -p_1 d_{\omega 3}& 0 &  p_3 d_{\omega 1}\\
      p_1 d_{\omega 2}& - p_2 d_{\omega 1}&0
  \end{array}\right)
  -a
   \left(\begin{array}{ccc}
     \rho_1 & \rho_1&\rho_1\\
     \rho_2&\rho_2&\rho_2\\
    \rho_3&\rho_3&\rho_3
  \end{array}\right),
  \eea
  where
   \be
  p_j=\frac{1 }{(\bm d_\omega\times \bm\rho)_j}, \quad \bm\rho=k\bm r, \quad a=\frac{3+2\rho^2-i\rho^3}{\rho^2(1+\rho ^2)}.
  \ee      
It follows from (\ref{matrix}) that the diagonal elements of matrix  $D_{lm}$ are dependent only on coordinates and do not depend on dipole components. Denote by $R_m$ and $I_m$  the real and imaginary parts of diagonal elements:
\be
\label{ReD}
R_m={\rm Re}\, D_{mm},\quad I_m={\rm Im}\, D_{mm}.
\ee
Let us calculate the quantities
\bea
\label{ReD2}
R&=&\left(\sum_m R_{m}^2\right)^{1/2}=\displaystyle\frac{3+2\rho}{\rho(1+\rho^2)},\\
\label{ImD}
I&=&\left(\sum_m I_{m}^2\right)^{1/2}=\displaystyle\frac{\rho^2}{1+\rho^2}.
\eea
Equations (\ref{ReD2}) and (\ref{ImD}) allow to find the distance  $r$ between the observer and the dipole, because  $R$ and $I$ are known quantities. For example, we get from  (\ref{ImD})
 \be
 \label{dist}
 r=\frac 1k\sqrt{\frac{I}{1-I}}.
 \ee
Coordinates of the unit vector $\bm n$ can be derived from equations  (\ref{matrix}),   (\ref{ReD}) and (\ref{ImD}):   
\be\label{direct}
n_m=-\frac{R_m}{R}=\frac{I_m}{I}.
\ee
As a consequence we obtain relations 
\be\label{Rel12}
IR_m=-RI_m,
\ee
which represent two independent  equations of  constraints for the diagonal elements of matrix $D_{lm}$. The third  constraint can be derived by elimination of variable $\rho$ from relations (\ref{ReD2}) and (\ref{ImD}). This gives 
\be
\label{Rel3}
R^2I=(3-I)^2(1-I) .
\ee
Hence, among the three complex diagonal elements $D_{lm}$, only three real numbers are independent.

Further we find the dipole vector $\bm d_\omega$. Substituting the found expression for $\bm\rho=\rho\bm n$  into   (\ref{matrix}) we get  (as  $n_m$ we take $-R_m/R$):
  \bea
  \label{matrixD}
 D_{lm}&=
  \left(\begin{array}{ccc}   
     0 & p_2 d_{\omega 3}& -  p_3 d_{\omega 2}\\
     -p_1 d_{\omega 3}& 0 &  p_3 d_{\omega 1}\\
      p_1 d_{\omega 2}& - p_2 d_{\omega 1}&0
  \end{array}\right)-\nn\\
  &-(R-{\rm i}I)
  \left(\begin{array}{ccc}   
     R_1 & R_1&R_1\\
     R_2&R_2&R_2\\
    R_3&R_3&R_3
  \end{array}\right),
  \eea
  where
   \be
  p_j=\frac{R }{\rho(\bm d_\omega\times \bm R)_j}.
  \ee           
   The left part of  (\ref{matrixD}) is known, the second matrix of the right side is already defined. Undefined remains only vector  $\bm d_\omega$. We introduce notation  $F_{lm}$ for the matrix
     \be
  \label{matrixF}
F_{lm}= \rho D_{lm}+\rho (R-{\rm i}I)
   \left(\begin{array}{ccc}    
     R_1 & R_1&R_1\\
     R_2&R_2&R_2\\
    R_3&R_3&R_3
  \end{array}\right).
  \ee
Then the vector $\bm d_\omega$ satisfies the matrix equation
     \be
  \label{dmatrix}
\rho   \left(\begin{array}{ccc}   
     0 & p_2 d_{\omega 3}& -  p_3 d_{\omega 2}\\
     -p_1 d_{\omega 3}& 0 &  p_3 d_{\omega 1}\\
      p_1 d_{\omega 2}& - p_2 d_{\omega 1}&0
  \end{array}\right)
=F_{lm}.
  \ee
 One can easily prove that the equations for diagonal elements are satisfied identically. For the off-diagonal elements we have six linear homogeneous equations:
 \begin{align}\label{eq1-3}
  \begin{split}
&F_{12}R_3d_{\omega 1}+(1-F_{12}R_1)d_{\omega 3}=0, \\
&F_{13}R_2d_{\omega 1}+(1-F_{13}R_1)d_{\omega 2}=0, \\
&F_{21}R_3d_{\omega 2}+(1-F_{21}R_2)d_{\omega 3}=0, 
\end{split}\\
\label{eq4-6}
\begin{split}
&(1-F_{23}R_2)d_{\omega 1}+F_{23}R_1d_{\omega 2}=0,\\
&(1-F_{31}R_3)d_{\omega 2}+F_{31}R_2d_{\omega 3}=0,\\
&(1-F_{32}R_3)d_{\omega 1}+F_{32}R_1d_{\omega 3}=0.
\end{split}
 \end{align} 
It follows directly from (\ref{dmatrix}) that $\sum_l R_lF_{lm}=1$. This defines next three constraints for elements  $D_{lm}$
 \bea\label{relat}
 \bsp
&R_2F_{21}+R_3F_{31}=1,\\
&R_1F_{12}+R_3F_{32}=1,\\
&R_1F_{13}+R_2F_{23}=1.
\end{split}
\eea
The last relations transform the equations (\ref{eq1-3}), into equations (\ref{eq4-6}) and vice versa. Hence, we can define the vector $\bm d_\omega$ for example, from   (\ref{eq1-3}), which by use of  (\ref{relat}) can be simplified as follows 
  \bea\label{eqd3}
  \bsp
&F_{21}d_{\omega 2}+F_{31}d_{\omega 3}=0,\\
&F_{12}d_{\omega 1}+F_{32}d_{\omega 3}=0,\\
&F_{13}d_{\omega 1 }+F_{23}d_{\omega 2}=0.
 \end{split}
\eea 
Here $F_{lm}$ and $d_{\omega l}$ are complex numbers. Substituting $F_{lm}$  as (see also (\ref{dcomp}))
$
F_{lm}=F_{lm}^0\exp {\rm i} \psi_{lm},
$
where $F^0_{lm}$ and $\psi_{lm}$ are real numbers, we re-arrange the last equations to a set of six real equations
\bea
\label{eqdR3}
F^0_{21}d_{02}- F^0_{31}d_{03}=0,\nn\\
F^0_{12}d_{01}- F^0_{32}d_{03}=0,\nn\\
F^0_{13}d_{01 }-F^0_{23}d_{02}=0,\\
\alpha_2-\alpha_3=\psi_{31}-\psi_{21}+(2n+1)\pi, \nn\\
\alpha_1-\alpha_3=\psi_{32}-\psi_{12}+(2n+1)\pi,\nn\\
\alpha_1-\alpha_2=\psi_{23}-\psi_{13}+(2n+1)\pi,\nn
\eea 
where $n$ is integer. The first three equations define the ratio between the Fourier transform of the dipole components, the next three equations assign the phase difference between the components of vector $\bm d_\omega$. All six equations define  the dipole polarization ellipse up to a scaling multiplier. There is no way to define this multiplier using only the coordinate derivatives of $\bm H_\omega$, because the initial matrix equation contains only the dipole components ratio. Absolute values of the dipole components or the scaling multiplier one can calculate using the Stokes parameter $s_0$. For example one can use  (\ref{mashtab}) and exclude the distance between the observer and the dipole by use of  (\ref{dist}).
\section{Alternative solutions}\label{sec4}
 The elements of matrix  $D_{nm}$ are not independent (if the source is in fact a point dipole). They are connected by nine real equations which can be written in a form of three real equations  (\ref{Rel12}) and  (\ref{Rel3}), and three complex equations  (\ref{relat}). Hence, only nine real numbers  out of 18 real values of matrix  $D_{nm}$ are independent. This number of independent variables is theoretically enough to solve the inverse problem, i.e. to find nine real numbers -- three components of the radius-vector and three complex numbers  $\bm d_\omega$ --  the  amplitudes and phases of the dipole.  Evidently, the solution of the inverse problem can be presented in different forms, but all of them can be transformed to each other by use of  equations of constraint. 

Further we show, for example, that the inverse problem can be solved  without recourse to the diagonal elements of matrix  $D_{nm}$.  It is easy to show that  (\ref{matrix}) yields 
\bea
\label{alt1}
\bsp
&D_{21}\rho_2+D_{31}\rho_3=1+a(\rho^2_2+\rho^2_3),\\
&D_{12}\rho_1+D_{32}\rho_3=1+a(\rho^2_1+\rho^2_3),\\
&D_{13}\rho_1+D_{23}\rho_2=1+a(\rho^2_1+\rho^2_2).
\end{split}
\eea
There are no dipole components in this set of equations. Hence, we can solve the set with respect to the radius-vector  $\bm r$. Substituting the solution for $\bm r$ back into   (\ref{matrix}), we can find the vector $\bm d_\omega$.

Another method, may be practically more useful, is the following. Suppose the vector $\bm H_\omega$  is not linearly polarized. Let the coordinate system be defined so that  $OZ$ is orthogonal to the plane of  polarization of vector  $\bm H_\omega$.  Then  $x=y=0$ and vector  $\bm\rho$ is presented by the only Cartesian component $\rho_3$. The sign of this component is  not defined so far. In this case only first two of  equations ($\ref{alt1}$) remain: 
\be
D_{31}\rho_3=1+a\rho^2_3, \quad D_{32}\rho_3=1+a\rho^2_3,
\ee
because the elements  $D_{13}$ and $D_{23}$ are undefined in accordance with   (\ref{matrix}).
The last relations yield $D_{31}=D_{32}$. Substituting  $a$, we get
\be\label{com-speed}
D_{31}\rho_3=D_{32}\rho_3=\frac{-3-\rho^2+{\rm i}\rho^3}{1+\rho^2}.
\ee
Real parts of  $D_{31}$ and $D_{32}$ define the rate of magnetic field amplitude as a function of distance. One can see from the last expression that the field amplitude is decreasing with the distance, no matter how the axis  $OZ$ is directed. This gives us direction from the observer to dipole which we did not defined until now. Let the axis  $OZ$ point from the dipole to observer. Then  $\rho_3>0$ and $\rho_3=\rho$. Let us find the rates of amplitude and phase by extracting real and imaginary parts from the last equation.
For the real part we obtain relations
\be\label{speed}
\frac{1}{k|H_{\omega 1}|}\frac{{\rm d} |H_{\omega 1}|}{{\rm d}r}=\frac{1}{k|H_{\omega 2}|}\frac{{\rm d} |H_{\omega 2}|}{{\rm d}r}= -\frac{2+\rho^2}{\rho(1+\rho^2)}.
\ee
 In the near zone, at the distances much less than the wave length ($r\ll k^{-1}, \, \rho\ll 1$) amplitude of magnetic field drops with the distance as  $-2/r$, but in far zone this rate is half as great. Equation  (\ref{speed}) is a cubic equation on $r$.
Similarly, the imaginary part of   (\ref{com-speed}) gives the rate of phase change $\varphi'$ with the distance
\be\label{phasSpeed}
\varphi'=\frac{{\rm d}\varphi_1}{k\,{\rm d}r}=\frac{{\rm d}\varphi_2}{k\,{\rm d}r}=\frac{\rho^2}{1+\rho^2}.
\ee
Looking at the equations (\ref{speed}) and  (\ref{phasSpeed}) we see that at small distances  ($r\ll k^{-1}$) the phase rate is much less than the relative amplitude rate. As to large distances, the phase rate is equal to  $k$ and does not depend on  $r$. Hence, the solution of the last equation with respect to the distance
\be\label{dist1}
r=\frac 1k\sqrt{\frac{\varphi'}{1-\varphi'}}.
\ee
can not be used, because at  large distance $\varphi'\to 1$.
It would be expected, because in the wave zone, at distances much greater than the wave length, the field is a field of radiation. Hence, the phase displacement at distance of a wave length is equal to $2\pi$ for any $r$.

\section{Uniqueness and stability of the solutions}\label{stab}
Uniqueness of the obtained solutions follows directly from  the used methods. Namely, we have found all solutions of a set of algebraic  equations. In some cases the solution is obtained in different forms like, for example, expressions for $r$ given by equations (\ref{ReD2}) and (\ref{ImD}).  
The first equation allows to find $r$ from the real part of diagonal elements of matrix  $D_{lm}$, while the second one uses the imaginary part of this matrix. It means that in the first case $r$ is calculated by the use of measurements of the gradient of magnetic field amplitude, while in the second case calculations use measurements of the phase gradient  $\partial \varphi_m/\partial x_m$ (see equation (\ref{MatD})). The real and imaginary parts of the field gradient are not independent because the nine partial derivatives (\ref{MatD}) are defined only by three components of the dipole (\ref{dcomp}). The real and imaginary parts of the diagonal elements are bound up with equation (\ref{Rel3}).
Hence, there can be different forms of the solution, but all of them should give the same result in numerical calculations. If in some real experimental  situation, solutions for $r$ derived from equations (\ref{ReD2}) and (\ref{ImD}) are different, then it can indicate that the field measured is not a field of a dipole. 
The same reasoning is applicable to any other results of this paper, because as mentioned above, the initial set of equations is  overdetermined. One do not need to know all the derivatives of the field in order to solve the inversion problem. 

Let us estimate the stability of the obtained solutions. We study how errors in the measurements of the magnetic field translate into errors in the reconstructions. We start with analysis of expressions for $r$ (\ref{ReD2}) and (\ref{ImD}), then we proceed to  stability of the solution for $\bm n$ and finally we discuss  stability of the  solution for the Fourier transform of the dipole vector.

Let the reconstructed parameter $y$ be a function of measurements $x$: $y=f(x)$. We think of small reconstruction error  $\delta y$ as the module of the linear part of change in $y$ corresponding to the small change $\delta x$  in $x$. So we write
\[\delta y=\left|\frac{{\rm d}f(x)}{{\rm d}x}\delta x\right |.\]

Taking derivative from formula  (\ref{ReD2}) we obtain the relative error for $\rho$
\be\label{delta-ro}
  \frac{\delta \rho}{\rho}=\eta\frac{\delta R}{R},\quad \eta=\frac{(1+\rho^2)(3+2\rho^2)}{(1+2\rho^2)(3+\rho^2)}.
\ee
It is easy to check that $\eta\leq 1$. Next, we express  $\delta R/R$ in terms of  $\delta R_m$:
\[
\frac{\delta R}{R}=\frac{R_1\delta R_1+R_2\delta R_2+R_3\delta R_3}{R_1^2+R_2^2+R_3^2}.
\]
Let $R_g$ be the greatest among $|R_1|, |R_2|, |R_3|$ and we assume that $\delta R_1=\delta R_2=\delta R_3=\delta R_g$. Then
\be\label{3delta}
\frac{\delta R}{R}\leq3\frac{\delta R_g}{|R_g|}.
\ee
According to definition (\ref{MatD})
\[
R_g=\frac{\partial |H_g|}{k|H_g|\partial x_g }\approx \frac{\Delta |H_g|}{k|H_g|\Delta x_g}\, ,
\]
where  $\Delta |H_g|=|H_g|_2-|H_g|_1$ and $\Delta x_g=(x_g)_2-(x_g)_1$ are the finite differences measured in order to calculate the gradient of the magnetic field. The relative error in $R_g$ is calculated then as
\be\label{dRkR}
\frac{\delta R_g}{|R_g|}=\frac{\delta(\Delta |H_g|)}{\Delta|H_g|}+\frac{\delta|H_g|}{|H_g|}+\frac{\delta|\Delta x_g|}{|\Delta x_g|}.
\ee
Combining this with the expressions (\ref{delta-ro}) and  (\ref{3delta}) we get the estimation of relative error in $\rho$
\be\label{rel1}
 \frac{\delta \rho}{\rho}\leq 3\eta\left(\frac{\delta(\Delta |H_g|)}{\Delta|H_g|}+\frac{\delta|H_g|}{|H_g|}+\frac{\delta|\Delta x_g|}{|\Delta x_g|}\right).
 \ee
 Hence, the relative error in $\rho$ of formula  (\ref{ReD2}) is linearly dependent on the sum of relative errors in: measurements of the magnetic field difference in two near points of space; measurement of the absolute value of the magnetic field; and measurement of the distance between the two near points. The inverse problem in this case is said to be a well-posed problem.
 
Now we make the some analysis with respect to the solution (\ref{dist}). Taking derivative from equation (\ref{dist}) we get
\[
\frac{\delta\rho}{\rho}=\frac{1}{2(1-I)}\frac{\delta I}{I}.
\]
The denominator of this equation shows that the relative error in $\rho$ grows infinitely as $I\to 1$ or, as it is seen from (\ref{ImD}), as $\rho\to\infty$. Substituting $\delta I/I$ from equations  (\ref{MatD}) and (\ref{ImD}) we obtain
\be\label{rel2}
  \frac{\delta \rho}{\rho}\leq \frac 32 (1+\rho^2)\left(\left|\frac{\delta(\Delta\varphi_g)}{\Delta\varphi_g}\right|+\left|\frac{\delta(\Delta x_g)}{\Delta x_g}\right|\right),
\ee
$g$ is the subscript of the greatest among $|\Delta\varphi_i/\Delta x_i|$. Again, the error in reconstructed data depends linear on the errors in measurement. But this time the constant of proportionality $1+\rho^2$ grows with the distance $r$. Fig. \ref{fige4} shows an example of such behaviour. Therefore, at great distances $r\gg\lambda$ expression  (\ref{ReD2}) gives better results than equations (\ref{ImD}) or (\ref{dist}). The same conclusion is applicable to the pair of solutions (\ref{speed}) and (\ref{dist1}).

Let us estimate errors in the unit vector of direction (\ref{direct}). Using a vector form $\bm R=(R_1,R_2,R_3)$ we obtain next expression for the differential of $\bm n$
\[
{\rm d}{\bm n}=\frac{((\bm R\times {\rm d}\bm R)\times\bm R)}{R^3},\quad |{\rm d}{\bm n}|=\frac{{\rm d}R}{R}\sin \phi,
\]
where $\phi$ is the angle between vectors $\bm R$ and ${\rm d}\bm R$. As ${\rm d}n_m\leq|{\rm d}{\bm n}|$ we estimate the relative error in direction from the observer to the source of the field as
\[
\delta n_m=\frac{\delta R_g}{|R_g|}
\]
with $\delta R_g/|R_g|$ defined by equation (\ref{dRkR}). Following the reasoning in paragraph next to equation (\ref{rel1}) we conclude that the solution for $\bm n$ is stable.

Evidently, the same analysis is applicable to calculation of the dipole polarization from equations (\ref{eqdR3}) or calculations of the Fourier components of the dipole. This results also in the linear dependence between the errors in measurements and errors in calculations of the sought quantities. 

Hence, we can establish that the inversion problem discussed in this paper is well-posed and the obtained solutions are unique.

It is interesting to see how well the obtained solutions do work at very high and very low frequencies. As one can see, the frequency does not appear explicitly  in the final formulas. It is included only in the reduced distance $\rho=\omega r/c$. The errors of calculations depend only on relative distance $r/\lambda$. Therefore, for fixed distance, working with low frequency means working close to the source of the field. There are no problems in calculation of the distance in case $r\ll\lambda$ by use of gradient of magnetic field. Because the gradient of the field and the field itself are relatively  great and can be easily measured, even in the case of stationary field.  But use for calculation the phase gradient can cause large errors, because  at small distances or by slowly varying field the phase gradient is small and tends to zero as $\rho\to 0$ or $\lambda\to\infty$ (see equation (\ref{phasSpeed})).

 At high frequencies problems can be encountered with measurements of the derivatives of the magnetic field. These measurements must be made in a small area which is much less than the wavelength. But in case of short wavelength the measurement accuracy can be restricted, for example, by dimensions of the used device or probe. The relative error $\delta(\Delta |H_g|)/\Delta|H_g|$ in equation (\ref{rel1}) can be intolerable great in this case.

Significant errors can arise in case of weak magnetic field. There are at least two origins of errors in this case: i) noise -- it must be sufficiently less then the measured field; ii) error $\delta H/H$ (second item in equation (\ref{rel1})) in measurement of weak field  can become rather great.
\section{An example}\label{examp}
Here we test the derived formulas in a numerical experiment. We define some numerical values of the Fourier components of a dipole at some point of space. Then we calculate the magnetic field produced by the dipole at another  point of space. We use the calculated values of the field as ``measured'' vales and apply the developed method for location of the dipole position.  Next we add a  random disturbance to the calculated field values and calculate the position of the dipole again in order to analyse the stability of the solution.
  
\begin{figure}[htbp]\center
\includegraphics[width=2.5in]{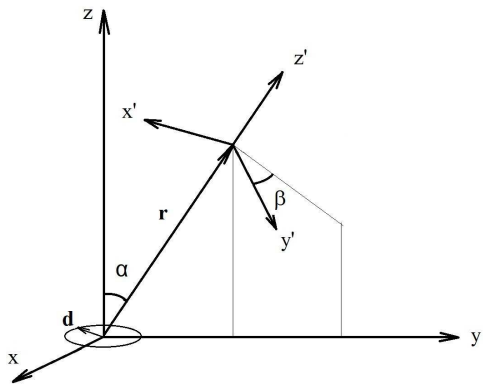}
\caption{Reference systems.}
\label{fig_1}
\end{figure}

Let the monochromatic component $\bm d_\omega e^{-{\rm i}\omega t}$ of  a dipole describe an ellipse in the coordinate plane $XY$ as shown in Fig. \ref{fig_1}. The dipole components are defined as follows:
\[
\bm d_\omega=(d_{\omega x}, d_{\omega y}, 0)=(d_x, {\rm i}d_y, 0),
\]
where $d_x$ and $d_y$ are real numbers.

  The field of the dipole is measured at a point with position vector $\bm r$. The magnetic field of the dipole changes in the plane orthogonal to the vector $\bm r$.
The observer adopts a coordinate system with axis $Z'$ orthogonal to the plane in which the magnetic field varies, and with arbitrary directed axes $X'$ and $Y'$. In order to calculate the magnetic field at point $\bm r$ we denote the angle between the axis $Y'$ and the plane $YZ$ by $\beta$.  Then the transition matrix from the coordinate system $(x, y, z)$ to the system  $(x', y', z')$  takes the form
\bea
T=
\begin{pmatrix}
\cos\beta&&-\sin\beta\cos\alpha&&\sin\beta\sin\alpha\\
-\sin\beta&&\cos\beta\cos\alpha&&-\cos\beta\sin\alpha\\
0&&\sin\alpha&&\sin\beta
\end{pmatrix}
\eea
Using equation (\ref{spectr}) we find the components of the magnetic field vector in the primed coordinate system
\bea
H_{\omega x'}=Ad_{\omega y'},\quad H_{\omega y'}=- Ad_{\omega x'},\quad H_{\omega z'}=0,\quad A=\frac{{\rm i} k}{r^2}({\rm i} kr-1){\rm e}^{{\rm i} kr}.
\eea
Here $d_{\omega j'}$, $j'=x', y'$ are the dipole components in the primed reference system, which are defined by equations
\[d_{\omega j'}=T_{jl}d_{\omega l},\]
where $T_{jl}$ are the elements of  transition matrix $T$.

In order to calculate the measurable values -- Fourier amplitude and phase --  we extract the real and imaginary parts of $H_{\omega j'}$. This gives
\bea\label{long}
H_{\omega x'}=\frac{k}{r^2}\left[Sd_x\sin \beta+Cd_y\cos\beta\cos\alpha-{\rm i}(Cd_x\sin\beta-Sd_y\cos\beta\cos\alpha)\right],\nn\\
H_{\omega y'}=\frac{k}{r^2}\left[Sd_x\cos \beta-Cd_y\sin\beta\cos\alpha-{\rm i}(Cd_x\cos\beta+Sd_y\sin\beta\cos\alpha)\right],\\
S=\sin\rho-\rho\cos\rho,\quad C=\cos\rho+\rho\sin\rho,\quad \rho=kr.\nn
\eea
We can simplify these expressions supposing that the observer chooses the axes $X'$ and $Y'$ in a way that the phase difference between $H_{\omega x'}$ and $H_{\omega y'}$ is $\pi/2$. One can see from the last equations that it happens if $\beta=0$ or $\pi/2$. Further we consider the case $\beta=0$. Then equations (\ref{long}) take the form
\bea
H_{\omega x'}&=&\frac{k}{r^2}d_y\cos\alpha(C+{\rm i}S),\nn\\
H_{\omega y'}&=&\frac{k}{r^2}d_x(S-{\rm i}C).\nn
\eea
The real values which are measured in practice are:  the amplitudes
\bea\label{ampl}
|H_{\omega x'}|=\frac{k^3}{\rho^2}d_y\cos\alpha\sqrt{1+\rho^2},\quad |H_{\omega y'}|=\frac{k^3}{\rho^2}d_x\sqrt{1+\rho^2},
\eea
and the phases
\bea\label{phases}
\varphi_{x'}=\arctan \frac{S}{C}\pm \pi n,\quad \varphi_{y'}=-\arctan \frac{C}{S}\pm \pi n,
\eea
$n$ are the integer numbers.

\begin{figure}[tbh!]
\centerline{\includegraphics[width=3in]{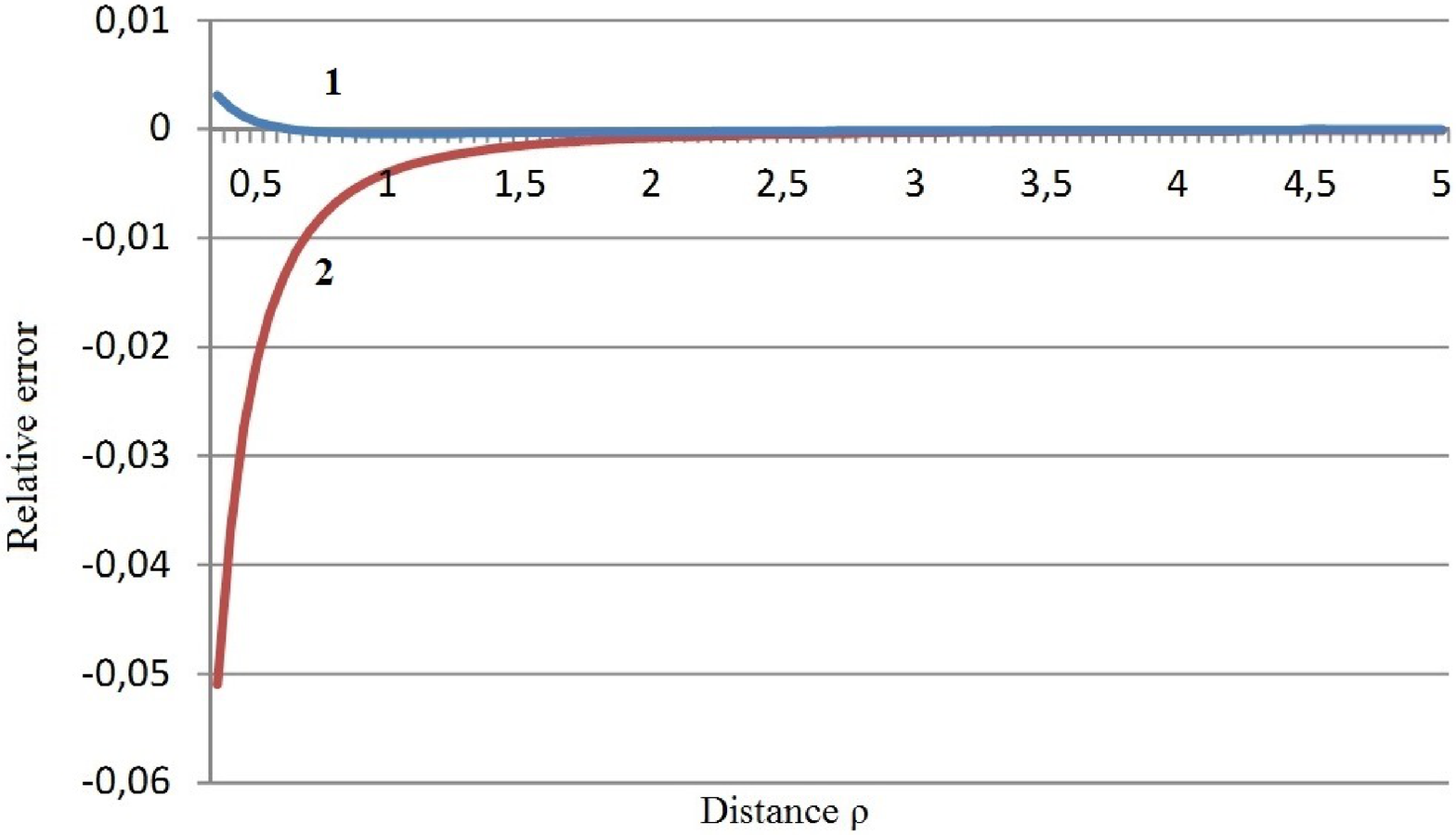}} \caption{Relative errors of the distance calculation. 1 -- error in the case when the distance is calculated from the formula  (\ref{dist1}); 2 -- error in the case when the distance is calculated from the formula  (\ref{speed}) } \label{fige2}
\end{figure}

Using equations (\ref{ampl}) and (\ref{phases}) we calculate amplitudes and phases at three close points along the axis $Z'$: at $z'=\rho$ and at $z'=\rho\pm\Delta \rho$. Then we calculate numerically the derivatives as
\bea\label{diff}
\frac{{\rm d}f(\rho)}{{\rm d}\rho}=\frac{f(\rho+\Delta\rho)-f(\rho-\Delta\rho)}{2\Delta\rho},
\eea
where
\bea
f=|H_{\omega x'}|,\, |H_{\omega y'}|,\, \varphi_{x'},\, \varphi_{y'}.\label{diff1}
\eea
These derivatives we use to calculate the distance between the observer and the source of the field according to formulas (\ref{speed}) and  (\ref{dist1}) and to find relative error of  these calculations. Calculated values of the derivatives of $|H_{\omega 1}|$ and $|H_{\omega 2}|$  in equation  (\ref{speed}) and derivatives of $\phi_1$ and $\phi_2$   equation  (\ref{phasSpeed}) are equal. The relative error $\varepsilon$ in calculation of $\rho$ is defined as follows: $\varepsilon=(\rho-\rho_0)/\rho_0$, where $\rho_0$ is the ``exact'' value of $\rho$ used by direct calculation of the magnetic field, and $\rho$ is the result of solution of the inverse problem.  The dependence $\varepsilon(\rho)$  is shown in Fig.~\ref{fige2}. In calculation of derivatives we put $\Delta\rho=0.05$. In other words, the field is ``measured'' at the points which are $\Delta r=0.05\lambda/2\pi$ apart.

Relative great errors at small distance, especially those represented by curve 2, are caused alone by numerical   differentiation. At small distances, when  $r$ is comparable with the wavelength $\lambda$, the phase, and especially the magnetic field amplitude grow rapidly. The  used method (\ref{diff}) for numerical differentiation is too rough in this case. Evidently, these errors can be made as small as desired by use of appropriate numerical methods of calculations.

\begin{figure}[tbh!]
\centerline{\includegraphics[width=3in]{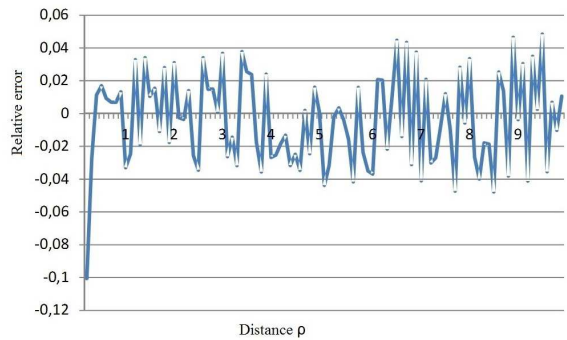}} \caption{Relative error which occur if the distance is calculated by use of equation (\ref{speed}). A 1\% noise is added.} \label{fige3}
\end{figure}

\begin{figure}[tbh!]
\centerline{\includegraphics[width=3in]{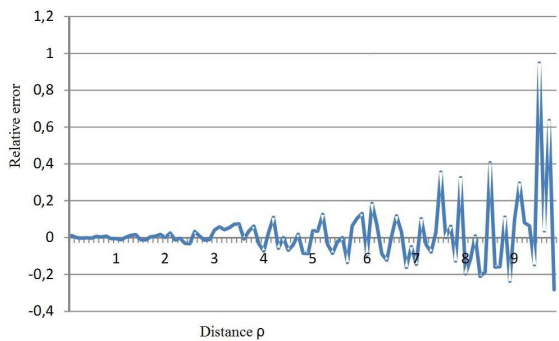}} \caption{Relative error which occur if the distance is calculated by use of equation (\ref{dist1}). A 1\% noise is added.} \label{fige4}
\end{figure}

In order to analyse the  stability of solutions  with regard to the noise, we have added to the ``measured'' data listed in  (\ref{diff1}) a 1\% random noise defined by formula $f_{\rm noise}=f(1+0.02 R)$, where $R$ is a random variable of interval $[-0.5, 0.5]$. The result is shown in Figs. \ref{fige3} and \ref{fige4}.
 As one can see, the solution  (\ref{speed}) is rather stable -- the relative error does not exceed 5\% above the distance $\rho=1$ ($r>\lambda/2\pi$). But the plot in Fig \ref{fige4}, which  represents the solution given by formula (\ref{dist1}) demonstrates the unacceptable errors at  large distances. These results of numerical modelling agree with the discussions following the equations (\ref{rel1}) and (\ref{rel2}).

\section{Conclusions and discussion}
%
       The inverse problem for a dipole magnetic field represented by its Fourier transform  can be solved in different ways, dependent on input data.  In general, unknown quantities are three coordinates of the field source, and three complex components of the dipole. This gives 9 unknown real quantities. Hence, at least 9 independent real numbers should be known to make the solution possible.  
 
If the magnetic field is not linearly polarized, information on its spectral properties (Fourier transform $\bm  H_\omega$ or Stokes parameters) makes it possible to define two opposite directions from the observer to dipole as a perpendicular to the plane in which the magnetic field vector is varying. Projection of the dipole polarization ellipse onto the plane orthogonal to this line is defined by  (\ref{delta}) up to a scaling multiplier. But the information on the spectral properties of the magnetic field is insufficient for calculation of distance between the dipole and observer. As additional input parameters derivatives of the magnetic field with respect to coordinates at the observation point can be used.

In Section \ref{sec3} solution is given for the case when all the nine derivatives from the magnetic field are known. 
 The inverse problem in this case has unambiguous solution.  Distance to the dipole can be calculated by formula (\ref{ReD2}), or (\ref{ImD}) or (\ref{dist}). Direction from the dipole to the observer is defined by the unit vector  (\ref{direct}), and the dipole polarization vector by  (\ref{eqdR3}). Absolute value of the dipole amplitude is given by equations  (\ref{Stokes3}) and  (\ref{mashtab}), because the distance  $r$ is already known at this stage. 

There are nine complex derivatives  $D_{lm}$ defined only by two vectors -- a real one $\bm r$ and a complex vector  $\bm d_\omega$. Hence, it can be no more than nine independent elements of matrix  $D_{lm}$. 
There are 9 real constraint equations for elements of this matrix which can be written down, for example, in the form (\ref {Rel12}), (\ref {Rel3}) and (\ref {relat}). 

In Section \ref{sec4} an example of inverse problem solution for a case when not all derivatives of the magnetic field are known is presented. It is shown that the solution of the inverse problem exists, if only non-diagonal elements of the matrix $D _ {lm} $ are known.The dipole coordinates  In this case are defined by (\ref {alt1}), and an ellipse of polarization of the dipole is described by equations (\ref {eqdR3}). If  polarization of the magnetic field is not linear and we are interested only in position of the field source, then it is sufficient to know the derivative of only one Cartesian coordinate of the magnetic field vector, which should be calculated along direction, orthogonal to the plane of magnetic field polarization. In this case the distance to a source is defined by equations (\ref {speed}) or (\ref {dist1}), and the direction to a field source is defined as a perpendicular to the plane of polarization of  vector $\bm H_\omega $.

Uniqueness and stability of the obtained solution are discussed in Section \ref{stab}. It is shown that the obtained solution of the inverse problem is unique and stable. But the relative errors depend on which data are  used for calculation of sought quantity. A numerical example in Section \ref{examp} demonstrates the theory developed in the previous sections.

\section*{Acknowledgments}

The authors would like to thank Dr. V.F. Gordeev and Dr I. Zhukova for valuable discussions. 
This work has been supported by the grant for LRSS, project No 224.2012.2 and by the Ministry of Education and Science of Russian Federation, project 14.B37.21.0774.

%
%

\end{document}